\documentclass[twocolumn, prl]{revtex4}

\usepackage{graphicx}
\usepackage{dcolumn}
\usepackage{bm}

\begin{document}

\title{Transference of Transport Anisotropy to Composite Fermions}

\date{\today}

\author{T.\ Gokmen}

\author{Medini\ Padmanabhan}

\author{M.\ Shayegan}

\affiliation{Department of Electrical Engineering, Princeton
University, Princeton, NJ 08544}

\begin{abstract}

When interacting two-dimensional electrons are placed in a large perpendicular magnetic field, to minimize their energy, they
capture an even number of flux quanta and create new particles called composite fermions (CFs). These complex electron-flux-bound
states offer an elegant explanation for the fractional quantum Hall effect. Furthermore, thanks to the flux attachment, the effective
field vanishes at a half-filled Landau level and CFs exhibit Fermi-liquid-like properties,
similar to their zero-field electron counterparts. However, being solely influenced by
interactions, CFs should possess no memory whatever of the electron parameters. Here we address a
fundamental question: Does an anisotropy of the electron effective mass and Fermi
surface (FS) survive composite fermionization? We measure the resistance
of CFs in AlAs quantum wells where electrons occupy an
elliptical FS with large eccentricity and anisotropic effective mass. Similar to their electron counterparts, CFs also exhibit
anisotropic transport, suggesting an anisotropy of CF effective mass and FS.

\end{abstract}


\maketitle

Clean two-dimensional (2D) electron systems provide one of most fertile grounds for
the observation of many-body phenomena in nature. When the 2D system
is cooled to low temperatures and placed in high, perpendicular
magnetic fields to minimize the electrons' kinetic energy, it
exhibits fascinating and often unexpected phenomena originating from
electron-electron interaction. Examples include the fractional
quantum Hall effect, electron Wigner crystallization, and the
formation of spin textures (Skyrmions) \cite{QHBook}. Two
decades ago the composite fermion (CF) picture was put forth, initially to describe the
fractional quantum Hall effect \cite{JainPRL89, HalperinPRB93, CFbook}. The CFs are formed
by attaching an even number of magnetic flux quanta to each electron
at high magnetic fields (Fig. 1). With this transformation, since
the attached flux exactly cancels out the external magnetic field at
half-filled Landau-levels (LLs), the CFs should ignore the very large external
magnetic field and behave as if they are at zero field. Such
remarkable behavior has indeed been confirmed in various experiments
\cite{QHbook, CFbook}. One particularly fundamental property of CFs
at exactly half-filled LLs is that they form a Fermi sea and
therefore possess a Fermi surface (FS) \cite{FootNote1}. This has also
been verified in several experiments \cite{QHBook, HalperinPRB93, CFbook}.

A natural question that arises is: In a
2D system with an anisotropic electron effective mass and FS (at zero
magnetic field), do the CFs retain such anisotropy? The answer to this question is
not obvious. One might argue that, since the CFs are a manifestation
of the electron-electron interaction, their physical properties
should only depend on the magnitude of the magnetic field which
quantifies this interaction, and not on the electrons' zero field ($B=0$)
properties. On the other hand, the interaction could be anisotropic
if the effective mass of electrons is anisotropic. In our study, we measure and compare the piezo-resistance of
electrons at $B=0$ and of CFs at LL filling factors
$\nu=1/2$ and $3/2$ in a 2D electron system confined to an AlAs
quantum well. In this system, the electrons can be re-distributed,
via the application of strain, between two conduction band valleys
each of which has an anisotropic FS. The piezo-resistance of CFs exhibits anisotropic transport, qualitatively similar to the
electrons' at $B=0$.

\begin{figure}
\centering
\includegraphics[scale=1]{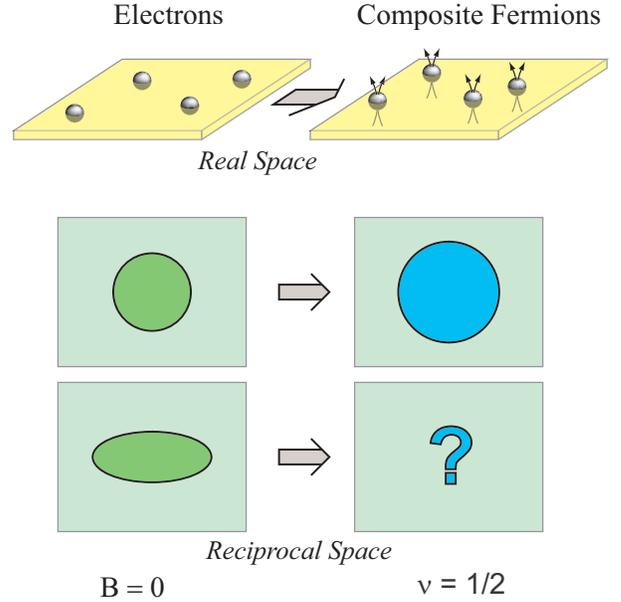}
\caption{Schematics of 2D electrons and composite fermions in real and reciprocal spaces.
Top panel: Schematics of 2D electrons and composite
fermions in real space. Lower panels: Schematics of isotropic and
anisotropic electron (left) and corresponding composite fermion
(right) Fermi surfaces.}
\end{figure}

We note at the outset that in a system with a parabolic energy vs. wave vector dispersion, such as AlAs electrons at $B=0$, the
anisotropies of the effective mass and the FS are linked. In a simple (Drude) model, the transport anisotropy in such a system is a
direct consequence of the effective mass anisotropy and the FS anisotropy which typically leads to an anisotropic scattering time. The
applicability of a parabolic dispersion or Drude model to CFs is not clear. We would like to emphasize, however, that our data provide
evidence that the transport anisotropy we observe is not merely a consequence of scattering time anisotropy of the CFs, but rather
points to an anisotropy of the CF mass and FS.

We studied 2D electrons confined to a 12 nm-wide AlAs
quantum well which was grown by molecular beam epitaxy on a (001)
GaAs substrate. In the absence of in-plane strain, the 2D electrons
in this system occupy two energetically degenerate conduction-band
valleys. The valleys are centered at the X-points of the Brillouin
zone, and have an anisotropic FS \cite{ShayeganPSS2006},
characterized by longitudinal and transverse effective masses,
$m_l=1.05$ and $m_t=0.205$ (measured in units of free electron
mass). The two valleys have their major axes along either the [100]
or the [010] crystal directions and we refer to these as X and Y
valleys, respectively. Their degeneracy can be lifted by applying
in-plane symmetry-breaking strain $\epsilon = \epsilon_{[100]} -
\epsilon_{[010]}$, where $\epsilon_{[100]}$ and $\epsilon_{[010]}$
are the strain values along [100] and [010]. The valley splitting
energy is given by $E_V = \epsilon E_2$ where $E_2$ is the
deformation potential which in AlAs has a band value of $5.8$ eV.
Positive strain pushes the energy of the X valley up relative to the
Y valley causing electrons to transfer from X to Y, and vice versa
for negative strain \cite{ShayeganPSS2006}. To apply tunable strain
we glued the sample to one side of a piezo-electric (piezo) stack
actuator and used a strain gauge glued to the other side to measure
the applied strain \cite{ShayeganPSS2006}. We studied a sample
patterned into a standard Hall bar mesa with its length along the
[100] crystal direction. Using a metal gate deposited on the
sample's surface we varied the 2D electron density $n$, between
$1.2$ and $4.8\times10^{11}$ cm$^{-2}$. The transport measurements
were performed down to a temperature of $0.3$ K, and up to a
magnetic field of $16$ T, using low-frequency lock-in techniques.

\begin{figure}
\centering
\includegraphics[scale=1]{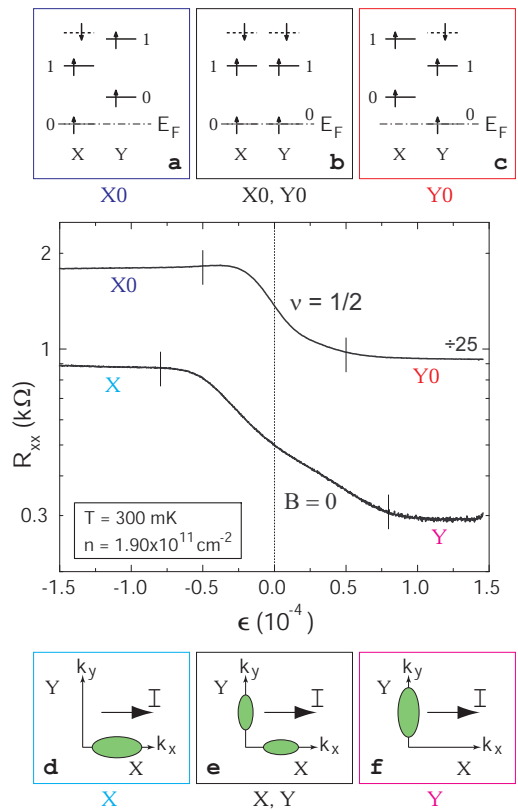}
\caption{The piezo-resistance data, valley occupations and Landau level diagrams of electrons and composite fermions at $\nu=1/2$.
Main: The piezo-resistance of electrons at $B=0$ and
composite fermions at $\nu=1/2$ ($B=15.7$ T) at density
$n=1.9\times10^{11}$ cm$^{-2}$. (a), (b) and (c) Energy level
diagrams and the position of the Fermi energy ($E_F$) for negative,
zero, and positive strain at $\nu=1/2$. (d), (e) and (f) Valley
occupations for negative, zero, and positive strain at zero magnetic
field.}
\end{figure}

An example of a piezo-resistance trace taken at $B=0$ is shown in
Fig. 2. The data is consistent with the conventional
piezo-resistance seen in multi-valley semiconductors
\cite{SmithPRev54,ShkolnikovAPL04}. The resistance drops for
positive values of strain as electrons are transferred to the Y
valley whose mobility along the current direction (the [100]
direction) is higher because of its smaller effective mass $m_t$
along this direction (Fig. 2(f)). For negative values of strain, the
resistance rises because now the electrons are transferred to the X
valley which has a large effective mass along [100] (Fig. 2(d)). At
large enough values of strain (either positive or negative) the
resistance saturates once one of the valleys is completely depleted
and inter-valley electron transfer stops. The saturation value of
the resistance when all electrons are in X valley is about a factor
of 3 larger than the resistance when electrons are in Y valley. This
is a direct consequence of the mass and FS anisotropy of the
valleys at zero field \cite{ShayeganPSS2006, Dorda78PRB, Ando82RevModPhys}.

Before discussing the $\nu = 1/2$ data shown in Fig. 2, we first
describe the energy level structure of the 2D system in a
perpendicular magnetic field. The magnetic field quantizes the
orbital motion of the electrons and forces them to occupy a discrete
set of energy levels (LLs) separated by the cyclotron energy, $\hbar
\omega_c = \hbar e B/m^*$, where $B$ is the magnetic field and
$m^*=\sqrt{m_l m_t}$ is the cyclotron effective mass. In our system,
there are four sets of these LLs, one for each spin and valley
combination. The energy splitting between oppositely polarized spin
levels is given by the Zeeman energy, while the levels corresponding
to different valleys are separated by $E_V$. We label each of the
energy levels according to their valley (X or Y), spin ($\uparrow$
or $\downarrow$) and LL index ($0, 1, ..$). However, for the density
range of our study, the parameters for AlAs electrons are such that
the Zeeman energy is larger than the cyclotron energy. Therefore,
all the LLs that are of concern here ($\nu < 2$) have the same spin.

The piezo-resistance trace in Fig. 2 taken at $\nu=1/2$
qualitatively shows a very similar behavior to the $B=0$ trace. The
resistance decreases (increases) towards positive (negative) values
of strain and at high enough strains it saturates. For very negative
values of strain, the $\nu=1/2$ CFs form in the lowest LL of the X
valley, namely X0 (Fig. 2(a)). For very positive values of strain,
the CFs are transferred to the lowest LL of the Y valley, i.e., Y0
(Fig. 2(c)). If the CF mass and FS were both isotropic, we would expect
to observe a symmetric piezo-resistance as the resistance should be
the same whether the CFs are either in X0 or Y0. The asymmetry of
the piezo-resistance at $\nu=1/2$ therefore indicates that the
CFs qualitatively retain the anisotropy of $B=0$ electrons.

\begin{figure}
\centering
\includegraphics[scale=1]{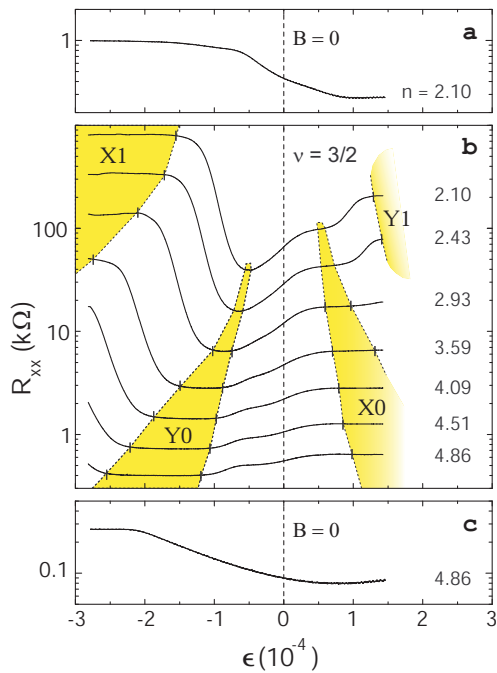}
\caption{The piezo-resistance data of electrons at $B=0$ and composite fermions at $\nu=3/2$.
(a) and (c) The piezo-resistance of electrons at $B=0$ at
densities $n=2.10$ and $4.86\times10^{11}$ cm$^{-2}$. (b) The
piezo-resistance of composite fermions at $\nu=3/2$ for several
densities, given in units of $10^{11}$ cm$^{-2}$. The traces are
offset vertically for clarity.}
\end{figure}

\begin{figure}
\centering
\includegraphics[scale=1]{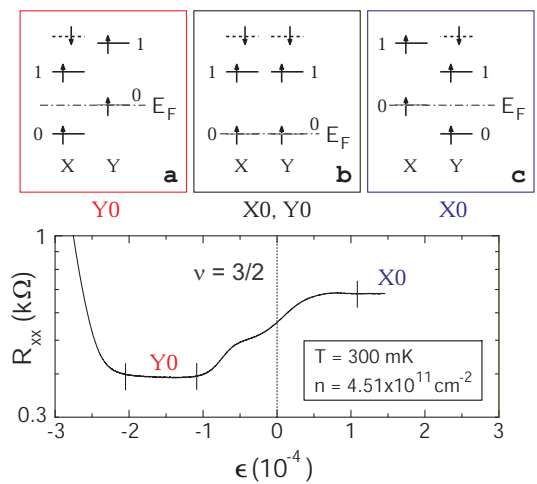}
\caption{The piezo-resistance of composite fermions at $\nu=3/2$ at high density.
Lower panel: The piezo-resistance of composite fermions at
$\nu=3/2$ at density $n=4.51\times10^{11}$ cm$^{-2}$. (a), (b) and
(c) Energy level diagrams for negative, zero, and positive strain at
$\nu=3/2$.}
\end{figure}

Although the piezo-resistance traces at $B=0$ and $\nu=1/2$ are
qualitatively very similar, there are two important quantitative
differences. First, as can be seen in Fig. 2, at $\nu=1/2$ the
resistance changes and saturates with strain more quickly compared
to $B=0$. This is because, as reported previously, the (valley
splitting) energy required to fully polarize the CFs is smaller than
the energy needed to valley polarize the electrons
\cite{BishopPRL07}. Second, in the density range of
$1.2-1.9\times10^{11}$ cm$^{-2}$ the resistance value when $\nu=1/2$
CFs are formed in X0 is only a factor of 1.5 to 2 larger than the
resistance value when CFs are formed in Y0. This resistance ratio of
CFs ($r_{CF}$) is about a factor of two smaller than the resistance
ratio of electrons ($r_e$) at $B=0$. We will return to this
difference later in the paper.

The piezo-resistance traces measured at $\nu=3/2$, while more
subtle, provide additional, strong evidence for the anisotropy of
CF mass and FS. Data taken for a range of densities are shown
in Fig. 3. The $\nu=3/2$ data reveal remarkably more features than
the $B=0$ or $\nu=1/2$ data. Instead of changing monotonically with
strain and showing only one transition, the piezo-traces at
$\nu=3/2$ exhibit a non-monotonic behavior with strain, suggesting
multiple transitions. The other striking feature of the
piezo-resistance at $\nu=3/2$ is that for small values of strain
(near zero) it shows the opposite trend compared to the $B=0$ and
$\nu=1/2$ data: The resistance increases with increasing strain,
instead of decreasing. As we will discuss below, all of these
features can be understood in the context of CFs with an anisotropic
mass and FS.

In Fig. 3, the $\nu=3/2$ piezo-resistance traces show four regions,
highlighted in yellow, where the resistance stays constant for
certain ranges of strain which depend on the 2D density. These
provide the key to understanding the data. The regions are labeled
by X0, Y0, X1, and Y1, indicating the energy level in which the
Fermi energy ($E_F$) resides. Note that the X0 and Y0 regions become narrower
and get closer to each other as the density is lowered. Also, given
the range of strain accessible in our experiments, we can observe
the regions X1 and Y1, where the resistance saturates, only at the
lowest densities.

In Fig. 4, we focus on the piezo-resistance trace at $\nu=3/2$ at a
density of $4.51\times10^{11}$ cm$^{-2}$. The opposite trend of the
piezo-resistance near zero strain compared to the $B=0$ or $\nu=1/2$
data is obvious. At $B=0$, positive strain pushes the X valley up in
energy and electrons are transferred from the X valley to the Y
valley. Similarly, at $\nu=1/2$ CFs are transferred from X0 to Y0
with positive strain. But as seen in the energy level diagrams of
Figs. 4(a) and (c), this situation is reversed for the case of CFs
formed at $\nu=3/2$. At $\nu=3/2$, for finite values of strain,
there is exactly one full LL and one half-filled LL. For small but
finite positive values of strain, the LLs of the X valley move up
compared to the Y valley, but $E_F$ stays at the X0 level
since this is the second lowest energy level (Fig. 4(c)). The Y0
level is of course completely full and therefore inert. Thus,
although the majority of the electrons are in the Y0 level, it is
the minority electrons in the X0 level which are at $E_F$
and should dominate the electrical transport at $\nu=3/2$. For
negative but small values of strain the X0 level becomes fully
occupied and inert and $E_F$ follows the Y0 level which
contains minority electrons (Fig. 4(a)). The opposite trends of the
piezo-resistance traces at $\nu=1/2$ and 3/2 (at small magnitudes of
strain) can therefore be understood by considering the position of $E_F$ and again assuming an anisotropic CF mass and FS.

Note that if the magnitude of strain is sufficiently small so that
the cyclotron energy is larger than the valley splitting energy, but
large enough so that the CFs are formed fully in the X0 or Y0
levels, we expect a region of constant resistance as seen and marked
by X0 and Y0 in Figs. 3 and 4 data. Interestingly, the ratio of the
resistance values for X0 and Y0 regions for $\nu=3/2$ traces is
between 1.5 and 2.3 in our available density range; this is similar
to the ratio $r_{CF}$ seen for the $\nu=1/2$ traces.

\begin{figure}
\centering
\includegraphics[scale=1]{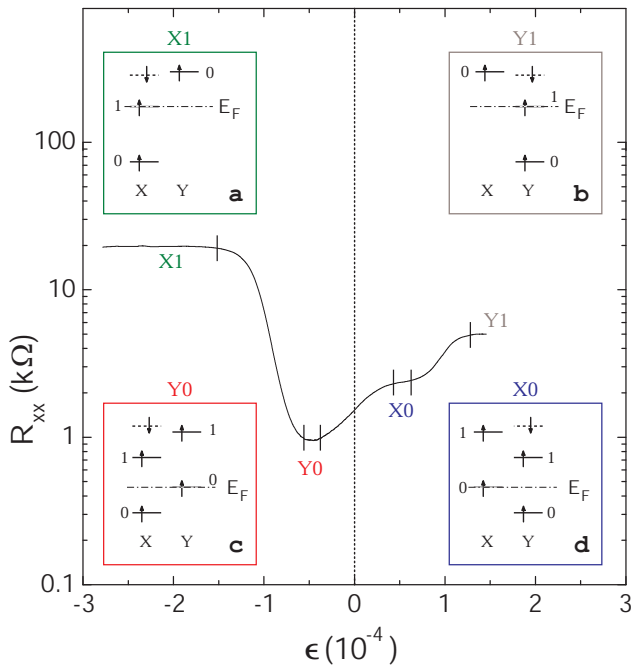}
\caption{The piezo-resistance of composite fermions at $\nu=3/2$ at
density $n=2.10\times10^{11}$ cm$^{-2}$. Insets show the four
possible energy level diagrams.}
\end{figure}

In Fig. 5 we demonstrate what happens at lower densities where we
can apply sufficient strain to transfer $all$ the electrons from one
valley to the other. This happens when the valley splitting energy
is larger than the cyclotron energy so that all the electrons occupy
the X valley (i.e., the X0 and X1 levels) or the Y valley (Y0 and
Y1). The $E_F$ then lies either at the X1 level for very
large negative values of strain (Fig. 5(a)), or at the Y1 level for
very large positive strains (Fig. 5(b)). The piezo-resistance trace
in Fig. 5 and other low-density traces in Fig. 3 reveal that, as the
electrons become fully valley polarized for either positive or
negative strains, the resistance increases. This is likely because
of the loss of screening upon full valley and spin polarization
\cite{GunawanNature07,VakiliPRL05}. Most remarkable, however, is
that the resistance rises much more for negative values of strain
compared to positive strains. The saturation value for very large
negative strains is in fact higher than the saturation value for
positive strains, similar to the $B=0$ and $\nu=1/2$ data. These
observations provide evidence for the CF transport anisotropy
even when the $\nu=3/2$ CFs are formed in the second LL.

As we mentioned before, the piezo-resistance at $B=0$ stems from the
anisotropy of the effective mass and FS. The FS anisotropy, however, typically leads to an anisotropic scattering time which can also
affect transport. In a simple Drude model, assuming isotropic scattering, the
resistance ratio ($r_e$) is equal to the mass ratio along the two
directions ($m_l/m_t$). For our system we measure $r_e \simeq 3$ which is smaller than the simple Drude prediction $m_l/m_t=5.1$.
This discrepancy can be understood by incorporating a scattering
time that is longer along the larger mass direction. This is a
reasonable assumption since along the large mass direction the Fermi
wave vector is larger and hence electrons should scatter
less because of their larger momentum \cite{Dorda78PRB, Ando82RevModPhys,TokuraPRB98}. Note that, in the most realistic scenario, the
anisotropic scattering time comes about because of the anisotropy of the FS and can only $reduce$ $r_e$ below the mass anisotropy
ratio.


The interpretation of the resistance ratio $r_{CF}$ for the CFs is
less clear. We emphasize that experimentally we measure a sizable $r_{CF}$
but $r_{CF}$ is always smaller than $r_e$. In a simple Drude model, this observation implies that either the mass anisotropy ratio for
CFs is smaller than for electrons at $B=0$ or the scattering for CFs is more anisotropic. The mass anisotropy ratio for CFs being
smaller than for electrons is plausible. In an ideal, isotropic 2D system, all the physical quantities of CFs are
determined by the Coulomb interaction ($\propto 1/\sqrt{x^2+y^2}$)
where $x$ and $y$ are components of the distance between two
electrons. Note that, at a fixed filling factor, this interaction is
solely quantified by the magnetic length $l_B=\sqrt{\hbar/eB}$
\cite{CFbook,JainPRL89,HalperinPRB93}. Now a system with an
anisotropic FS at $B=0$ can be mapped to a system with
isotropic FS at $B=0$ and an anisotropic Coulomb
interaction ($\propto 1/\sqrt{x^2\gamma^2+y^2/\gamma^2}$), where
$\gamma=(m_l/m_t)^{1/4}$. Obviously, in such
a case the strength of the Coulomb interaction depends not only on
$l_B$ but also on the direction and $\gamma$. If one assumes that
the transport mass of CFs along some direction is determined by the
strength of the Coulomb interaction along that direction, then the
mass anisotropy ratio of CFs is given by $\gamma^2=\sqrt{m_l/m_t}$
rather than $m_l/m_t$, consistent with the observation that $r_{CF}
< r_e$.

A theoretical study \cite{BalagurovPRB2000} that takes into account the mass anisotropy of electrons at $B=0$ predicts that the form of
the FS for CFs is identical to the zero field FS but that, despite this anisotropy, the CF effective mass is almost isotropic. It is
not obvious what this theory would predict for the resistance of the CFs. If the CF scattering time followed the anisotropy of its FS,
however, in a simple Drude model an isotropic CF mass would imply an $r_{CF}$ which has the opposite behavior to what we observe
experimentally. On the other hand, our data do not rule out a hypothetical situation where the effective mass is anisotropic but the FS
is isotropic.

Can the anisotropic transport we observe for CFs come only from anisotropic scattering? For example, it is known that disorder-induced
density variations at $B=0$ are accompanied by fluctuations in the effective magnetic field of CFs, resulting in more pronounced
scattering \cite{HalperinPRB93}. If such variations have a preferred crystal direction (e.g., along the strain axis) and their
magnitude increases with strain, then they could lead to anisotropic transport of CFs. Our data, however, provide strong evidence
against such scenario. First, the piezo-resistance we observe at $\nu=1/2$ and $3/2$ is clearly linked to the valley occupation of CFs
and not simply the magnitude of strain: The resistance only changes while the valleys are partially occupied and it saturates once the
CFs are fully valley polarized. Second, for small values of strain near zero, the $\nu=3/2$ piezo-resistance shows the opposite trend
compared to the $\nu=1/2$ case. The CF piezo-resistance anisotropy we observe therefore cannot result from a fixed anisotropic
scattering in a preferred direction, and is most likely related to the effective mass and FS anisotropy of CFs.

The piezo-resistance results presented here demonstrate
that, in a 2D electron system with an anisotropic effective mass and FS at
zero magnetic field, the CFs at filling factors 1/2 and 3/2 also
exhibit anisotropic transport, consistent with a qualitative transference of the electron mass and FS anisotropy to the CFs. Better
theoretical treatment is clearly needed to describe the CFs in an anisotropic system and
in particular quantitatively determine their resistance and mass or FS anisotropy. Also helpful would be measurements that directly
probe the
FS and its anisotropy, such as commensurability
oscillations in 2D systems with a periodically modulated electron
density \cite{SmetPRL98,GunawanPRL04}.

We thank the NSF and DOE for support, and J.K. Jain and B.I. Halperin for illuminating discussions.

\break

\end{document}